\documentclass[pra,aps,showpacs,twocolumn,preprintnumbers,superscriptaddress]{revtex4}
\usepackage{graphicx,bm,amssymb,amsmath}

\DeclareMathOperator{\Tr}{\,{\rm Tr}\,}                   

\begin{document}

\preprint{ITP-UU-08/56}

\title{A strongly interacting Bose gas: Nozi\`{e}res and Schmitt-Rink theory and beyond}
\author{Arnaud Koetsier}
\email{a.o.koetsier@uu.nl}
\affiliation{Institute for Theoretical Physics, Utrecht University, Leuvenlaan 4, 3584 CE Utrecht, The Netherlands}
\author{P. Massignan}
\affiliation{Institute for Theoretical Physics, Utrecht University, Leuvenlaan 4, 3584 CE Utrecht, The Netherlands}
\affiliation{ICFO-Institut de Ci\`encies Fot\`oniques, Parc Mediterrani de la Tecnologia, E-08860 Castelldefels (Barcelona), Spain}
\author{R.~A. Duine}
\affiliation{Institute for Theoretical Physics, Utrecht University, Leuvenlaan 4, 3584 CE Utrecht, The Netherlands}
\author{H.~T.~C. Stoof}
\affiliation{Institute for Theoretical Physics, Utrecht University, Leuvenlaan 4, 3584 CE Utrecht, The Netherlands}

\date{\today}

\begin{abstract}
We calculate the critical temperature for Bose-Einstein condensation in a gas of bosonic atoms across a Feshbach resonance, and show how medium effects at negative scattering lengths give rise to pairs reminiscent of the ones responsible for fermionic superfluidity. We find that the formation of pairs leads to a large suppression of the critical temperature. Within the formalism developed by Nozi\`{e}res and Schmitt-Rink the gas appears mechanically stable throughout the entire crossover region, but when interactions between pairs are taken into account we show that the gas becomes unstable close to the critical temperature. We discuss prospects of observing these effects in a gas of ultracold \textsuperscript{133}Cs atoms where recent measurements indicate that the gas may be sufficiently long-lived to explore the many-body physics around a Feshbach resonance.
\end{abstract}


\maketitle

\section{Introduction.}
In a Fermi gas there exists a smooth crossover connecting two apparently distant states, namely, the Bardeen-Cooper-Schrieffer (BCS) state which is found for weakly attractive interactions and the Bose-Einstein condensed (BEC) state of diatomic molecules obtained for weakly repulsive effective interactions~\cite{EaglesLeggetBECBCS}. This crossover between a superfluid of loosely bound pairs of fermions and a condensate of tightly bound dimers takes place in the vicinity of a Feshbach resonance, where the effective interaction changes from attractive to repulsive through a divergence of the coupling constant. What lies at the heart of this continuous crossover is the fact that a pair of fermions makes up a boson which is bound either by two-body effects in the BEC limit or by many-body effects in the BCS limit. It is this composite boson that undergoes Bose-Einstein condensation.

While fermions can condense only in the form of pairs, bosons can condense solitarily. A Bose gas at low temperatures is therefore subject to a competition between the condensation of atoms and of pairs. Pairing in a bosonic gas leads either to the formation of tightly bound molecules which are stable even in the vacuum, or to the creation of loosely bound pairs stabilized by the medium. The latter are reminiscent of Cooper pairs arising in BCS superconductivity.  In this article we investigate the properties of a Bose gas near a Feshbach resonance and address important issues regarding stability of such a strongly interacting Bose gas.

The BEC-BCS crossover for fermions has been recently explored in a series of ground breaking experiments with ultracold alkaline gases~\cite{BECBCSexperiments} by exploiting so-called Feshbach resonances which enable the interaction strength to be tuned. By contrast, early experimental attempts to create strongly interacting bosonic gases have been plagued by large losses~\cite{Ketterle03Sodium}. This is primarily because inelastic collisions provoking the decay of pairs into deeply bound states increase rapidly in the neighborhood of a resonance of the $s$-wave scattering length $a$. The decay imparts a kinetic energy to the products that is in general much larger than the confining potential, causing atoms to escape from the trap. Moreover, attractive atomic interactions may induce a mechanical instability at sufficiently large densities resulting in the collapse of the gas~\cite{Stoof94}. Both problems do not arise with fermions due to the stabilizing effects of Pauli blocking~\cite{Petrov04PRL}.

However, in recent experiments a gas of 6$s$ molecules composed of bosonic \textsuperscript{133}Cs atoms was observed to have relatively small inelastic losses~\cite{Ferlaino08,Knoop09}. This stability against decay offers the intriguing possibility of experimentally realizing a molecular BEC with bosonic atoms and of studying the BEC-BCS crossover in a bosonic gas.

The paper is organized as follows. In Sec.~\ref{sec:univ} we demonstrate that a crossover exists in the case of bosons that is analogous to the BEC-BCS crossover occurring in Fermi gases. We show that the compressibility obtained within the
Nozi\`{e}res and Schmitt-Rink (NSR) formalism~\cite{Nosieres85}  is positive throughout the normal phase, indicating that the gas may be mechanically stable. However, attractive interactions beyond the NSR formalism between long-lived preformed pairs of bosons can mechanically destabilize the gas and the elucidation of this effect is the subject of Sec.~\ref{sec:beyond}. In Sec.~\ref{sec:Cs} we discuss the experimental feasibility of observing such a crossover in  \textsuperscript{133}Cs, and present our conclusions in Sec.~\ref{sec:end}.

\section{Universal phase diagram}
\label{sec:univ}
In principle, a two-channel model is required to completely describe a Feshbach resonance~\cite{DuinePhD}. However, sufficiently close to the resonance, a single-channel description which lends itself to the NSR formalism suffices. The phase diagram describing the transition from the normal state of a bosonic gas with density $n$ to the condensed state of either atoms or pairs is then universal, in the sense that it depends only on a single interaction parameter, namely, the total scattering length $a$ divided by the average interparticle spacing $n^{-1/3}$. In the limit $n^{1/3}a\rightarrow 0^-$, atoms condense at the critical temperature for an ideal gas $T_a = (2\pi\hbar^2/mk_{\rm B})[n/\zeta(3/2)]^{2/3}$, where $n$ is the total atomic density. In the opposite limit $n^{1/3}a\rightarrow 0^+$ there exists a deep two-body bound state, and atoms bind into dimers which condense at the lower temperature $T_{\rm m} = T_a/2^{5/3}$. Within a simple two-body approximation (dashed line in Fig.~\ref{fig1}), we find a critical temperature for pair condensation $T_c$ that quickly interpolates between these two limits. However, within the NSR many-body approach (solid line in Fig.~\ref{fig1}) we find that $T_c$ is strongly suppressed on the attractive side of the resonance. As we will see in the following, this suppression is due to the population of a low-energy resonant two-body state brought about by many-body effects. This is in many ways reminiscent of what occurs in superconductors, where electrons in a medium pair up below a critical temperature in the presence of an attractive interaction. While in the fermionic case the critical temperature falls to zero exponentially, we find that here $T_c$ reaches its limiting value $T_a$ according to a power law~\cite{endnote1}. Within the NSR approach, the gas is found to be stable in the normal phase at all temperatures down to the critical temperature $T_c$. In Sec.~\ref{sec:beyond} we will see that inclusion of molecule-molecule interactions leads to a critical temperature that is higher than the NSR result, and in the region close to the resonance the gas acquires a negative compressibility at temperatures close to $T_c$.

\subsection{NSR calculation}
\label{sec:nsr}
\begin{figure}
\begin{center}
\includegraphics[width=\columnwidth]{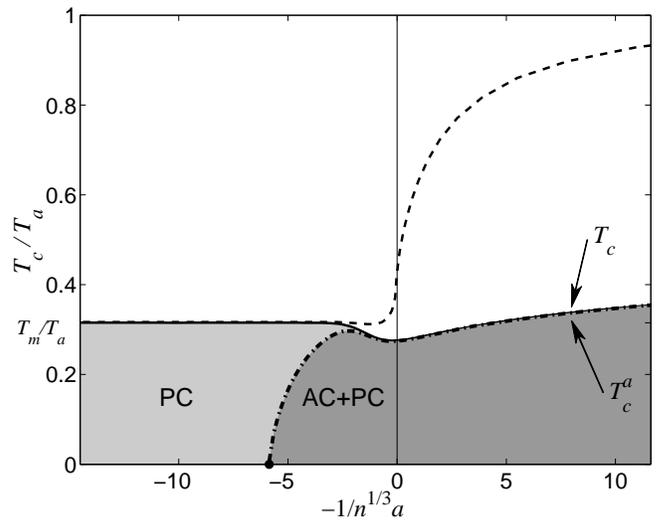}
\bf \caption{ \rm Universal phase diagram for bosons as a function of the interaction parameter
$1/n^{1/3}a$ where $n$ is the total density of atoms.
The solid line shows $T_c$ calculated within the NSR formalism, while the dashed line is the two-body result. The dash-dotted line is the critical temperature $T_c^a$ for the condensation of unpaired bosons in units of the critical temperature $T_a$ of an ideal atomic gas. The phase with a pair condensate (\textsc{PC}), which lies between $T_c$ and $T_c^a$, extends all the way to $-1/n^{1/3}a=+\infty$. Both an atomic condensate and a pair condensate (AC+PC) exist below $T_c^a$.\label{fig1}}
\end{center}
\end{figure}

The results from the NSR formalism are summarized in Fig.~\ref{fig1} which shows the critical temperatures $T_c$ and $T_c^a$ for the \mbox{Bose-Einstein} condensation of paired and unpaired bosons, respectively.
The critical temperature and the pressure of the gas was calculated within the grand canonical ensemble by introducing the thermodynamic potential $\Omega$ of a gas of particles of mass $m$ at temperature $T$. The pressure and the density of the gas are related to $\Omega$ by $P=-\Omega/ V$ and $n = -(1/V)\partial \Omega /\partial \mu$ respectively, where $V$ is the volume and $\mu$ the chemical potential. Following NSR, we write $\Omega$ as~\cite{Nosieres85}
\begin{equation}
\Omega = \frac{1}{\beta} \sum_{\bf k}\ln [ 1 - \exp(-\beta \xi_{\bf k})]
       + \frac{1}{\beta} \sum_{n,{\bf k}} \ln\frac{T^{\rm 2B}(0)}{T({\bf k},i\omega_n)}.
       \label{Omega}
\end{equation}
The first term is the contribution of an ideal gas of atoms with $\xi_{{\bf k}} = \hbar^2{\bf k}^2/2m -\mu$ and $\beta = 1/k_{\rm B} T$. The second term (rederived in Sec.~\ref{sec:beyond}) represents the contribution of paired atoms where $\omega_n=2\pi n/\hbar\beta$ are the even Matsubara frequencies and the two-body $T$-matrix is given by
\begin{equation}
    T^{\rm 2B}(z)=\frac{4\pi \hbar^2 a}{m}\frac{1}{1- a\sqrt{-zm/\hbar^2}},
    \label{T2B}
\end{equation}
where $z$ is the energy in the centre-of-mass frame. The instability towards pair condensation is signaled by the appearance of a pole in the $T$-matrix at ${\bf k}={\bf 0}$ and $i\omega_n=0$, which is the Thouless criterion~\cite{Thouless}. Neglecting many-body effects in the first instance, we take $T({\bf k},i\omega_n) = T^{\rm 2B}(z)$ with $z \equiv z({\bf k},i\omega_n)=i\hbar\omega_n +2\mu- \hbar^2 {\bf k}^2/4m$. This two-body result is plotted as a dashed line in Fig.~\ref{fig1}.

To include many-body effects, we take $T({\bf k},i\omega_n) = T^{\rm MB}({\bf k},i\omega_n)$, where the many-body T-matrix is~\cite{StoofNIST}
\begin{equation}
T^{\rm MB}({\bf k},i\omega_n)=\frac{1}{[T^{\rm 2B}(z)]^{-1}-\Xi({\bf k},i\omega_n)}. \label{TMB}
\end{equation}
Here, the renormalized correlation function in the particle-particle channel $\Xi({\bf k},i\omega_n)$ is given by
\begin{equation}
  \Xi({\bf k},i\omega_n) = \frac{1}{V}\sum_{{\bf q}}
    \frac{N( \xi_{{\bf k}/2+{\bf q}}) +N( \xi_{{\bf k}/2-{\bf q}}) }
                {i\hbar\omega_n -\xi_{{\bf k}/2-{\bf q}} -\xi_{{\bf k}/2+{\bf q}}},
\label{Xi}
\end{equation}
with the Bose factor $N(x) = (e^{\beta x} -1)^{-1}$. Within this formalism, the Thouless criterion for condensation yields the solid line in Fig.~\ref{fig1}.

\begin{figure}
\begin{center}
\includegraphics[width=0.95\columnwidth]{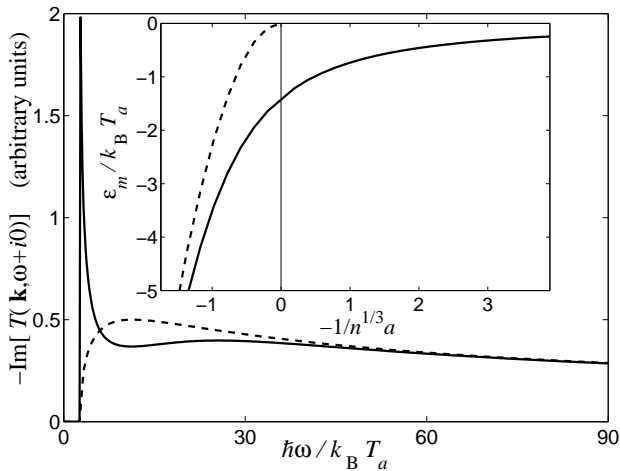}
\bf \caption{ \rm Spectral function of a pair evaluated at
$1/n^{1/3}a=2.18$, $T=T_c$ and $|{\bf k}| \simeq 0$. The two-body (dashed) and many-body (solid) calculations agree at high energy, but differ substantially at low energy. Also, a delta-function is present at positive frequency just below the continuum in the many-body case (not shown). At $|{\bf k}| = 0$ this delta function corresponds to the Thouless pole at $\omega=0$. Inset: binding energy of a pair at $T=T_c$. The dashed line is the ideal gas limit $\varepsilon_m=-\hbar^2/m a^2$, while the solid line is the NSR result given by $\varepsilon_m=2\mu$. \label{fig2}}
\end{center}
\end{figure}

The strong suppression of $T_c$ for negative scattering lengths can be understood from the spectral function for the pairs which is proportional to $\mathrm{Im}[T({\bf k},\omega+i0)]$ and is plotted in Fig.~\ref{fig2}. The many-body spectral function (solid line) acquires a narrow peak near zero energy in the continuum. For small momenta, a delta-function is also present just below the continuum that is associated with the presence of infinite-lifetime pairs (not shown). Both features are notably absent from the two-body spectral function (dashed line)~\cite{FalcoDuineStoof,DuinePhD}, and enhance the density of states for pairs at low energy leading to a reduction of $T_c$ compared to the two-body case. In addition, populating these many-body resonances reduces the atomic density thereby reducing the atomic condensation temperature $T_c^a$. The inset shows the binding energy at $T_c$ of the two-body bound state in the presence of the medium (solid line) which approaches the two-body result (dashed line) as $a\rightarrow0^+$. For $a < 0$ there is no bound state in the vacuum, and the binding energy in the medium is that of the pair associated with the delta-function in the spectral function.

\begin{figure}
\begin{center}
\includegraphics[width=0.93\columnwidth]{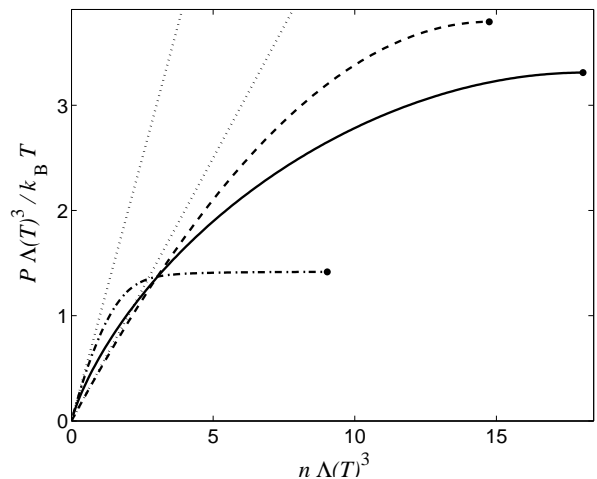}
\bf \caption{ \rm The pressure of the gas in its normal phase as a function of density in the NSR formalism for $-1/n^{1/3}a\simeq-10^3$ (dashed), $0$ (solid), and $10^3$ (dashed-dotted). Here, $\Lambda(T)=\sqrt{2\pi\hbar^2/mk_BT}$ is the thermal de Broglie wavelength.
The dotted lines are the ideal gas law for atoms and molecules. \label{figP1}}
\end{center}
\end{figure}

The pressure of the gas in the normal phase is plotted in Fig.~\ref{figP1} as a function of density up to the critical density for pair condensation which is denoted by a dot terminating the lines. We show results for $n^{1/3}a \simeq 0^+$, at resonance, and for $n^{1/3}a \simeq 0^-$. As expected, in the low density limit the first curve reproduces the ideal gas law for atoms ($P=n k_B T$), while the last yields the corresponding relation for molecules ($P=nk_B T/2$). We see that the compressibility ($=n^{-1}\partial n/\partial P$) calculated within the NSR approach is always positive, implying that the gas in its normal phase remains mechanically stable throughout the whole crossover.

Below $T_c$, bosons initially condense in the form of pairs~\cite{EvansRashid} and this region is denoted by PC in the phase diagram. Interactions between pairs are neglected within the \textsc{NSR} formalism, and below $T_c$ we therefore have an ideal Bose-Einstein condensate of pairs. The Thouless criterion is then always satisfied but the total number of atoms found from Eq.~(\ref{Omega}) is too small. This is because the number equation misses a nonzero pair-condensate contribution $n_c$ which emerges as a result of a zero-momentum divergence connected to the Thouless pole of the many-body $T$-matrix. For $T < T_c$ it is
\begin{equation}
  n_c = \hbar\frac{N(0)}{V}
  \frac{\partial T^{\rm MB}({\bf 0},0)}{\partial \mu}
  \left(\frac{\partial T^{\rm MB}({\bf 0},0)}{\partial i\omega_n}\right)^{-1}.
  \label{nc}
\end{equation}
The pair condensate is characterized by the BCS order parameter $\Delta_0\equiv\langle\Delta({\bf x},\tau)\rangle = V_0\langle\psi({\bf x},\tau) \psi({\bf x},\tau)\rangle \neq 0$ and $\langle\psi({\bf x},\tau) \rangle = 0$. Here $\Delta({\bf x},\tau)$ is the pairing field, $V_0$ is the bare atomic interaction and $\psi({\bf x},\tau)$ is the atomic field. Noting that the Fourier transform of the BCS Green's function $G_\Delta({\bf x},\tau;{\bf x}',\tau') = -\langle\Delta({\bf x},\tau)\Delta^*({\bf x}',\tau')\rangle$ can be written in terms of the many-body $T$-matrix as $G_\Delta({\bf k},i\omega_n) = 2\hbar T^{\rm MB}({\bf k},i\omega_n)$~\cite{Stoof94}, we find an expression for the order parameter $|\Delta_0|^2 = -\sum_{n} G_\Delta({\bf 0},i\omega_n) / \hbar\beta V$ in terms of the many-body $T$-matrix
\begin{align}
    |\Delta_0|^2 =&
     \frac{2\hbar N(0)}{V}
    \left[
    \frac{\partial}{\partial i\omega_n}\frac{1}{T^{\rm MB}({\bf 0},0)}
    \right]^{-1}.
    \label{delta0}
\end{align}
Now, by comparing the divergences in Eqs.~(\ref{nc}) and (\ref{delta0}) we express the gap in terms of the condensate fraction as
\begin{equation}
  |\Delta_0|^2 =2n_c
      \left[
    \frac{\partial}{\partial \mu}\frac{1}{T^{\rm MB}({\bf 0},0)}
    \right]^{-1}.
\end{equation}
Furthermore, the atomic dispersion in the pair condensate becomes a gapped Bogoliubov quasiparticle dispersion $\hbar\omega_{{\bf k}} = \sqrt{\xi_{{\bf k}}^2 - |\Delta_0|^2}$. Thus, condensation of atoms, signaled by $\langle\psi({\bf x},\tau)\rangle$ becoming nonzero, takes place when the quasiparticle dispersion vanishes at zero momentum. The onset of atomic condensation is hence found as the temperature $T_c^a$ at which $|\Delta_0| = -\mu$. This is the dash-dotted line in Fig.~\ref{fig1}.

It is only at the temperature $T_c^a < T_c$ that an atomic condensate (AC) of solitary bosons appears, and coexists with the condensate of pairs. The phase transition separating the AC and AC+PC regions becomes an Ising-like quantum phase transition when $T_c^a=0$~\cite{Radzihovsky04,Romans04}. We find no point at which the three phases (normal, PC, AC+PC) meet, as $T_c^a$ remains strictly smaller than $T_c$ at any value of the scattering length.

\subsection{Beyond NSR}
\label{sec:beyond}
Interactions between infinite-lifetime dimers corresponding to the positive frequency pole of the many-body $T$-matrix are neglected in the NSR theory presented above. In order to estimate the effect of these interactions in the normal phase, we add a mean-field shift $n_{\rm m} T_{\rm mm}$ to the binding energy of these infinite-lifetime dimers, and the corresponding additional interaction energy $V n_{\rm m}^2 T_{\rm mm}$ in the thermodynamic potential. Here, $T_{\rm mm} = 4\pi\hbar^2 a_{\rm mm}/2m$ is the molecular $T$-matrix, $a_{\rm mm}$ is the dimer-dimer $s$-wave scattering length and $n_{\rm m}$ is the density of dimers that is the contribution of the pole in the $T$-matrix to the number equation.

To calculate the dimer-dimer scattering length, we begin by performing a Hubbard-Stratonovich transformation and subsequently integrate out the atomic fields to obtain an effective action in terms of the pairing field
\begin{multline}
  S^{\rm eff}[\Delta,\Delta^*] = \frac{-1}{2V_0}\int_0^{\hbar\beta}\mkern-5mu d\tau \int\mkern-5mu d{\bf x} |\Delta({\bf x},\tau)|^2
  \\
  +\frac{\hbar}{2}\Tr\left[\ln(-{\bf G}_0^{-1})\right] - \frac{\hbar}{2}\sum_{n=1}^\infty\frac{1}{n}\Tr\left[({\bf G}_0{\bf\Sigma})^n\right],
  \label{eq:Seff}
\end{multline}
where the noninteracting Green's function is a $2\times 2$ matrix in Nambu space given by
\begin{multline}
  \hbar{\bf G}_0({\bf x},\tau;{\bf x}',\tau')^{-1} = \\
  \begin{pmatrix}
-\hbar \partial_\tau + \frac{\hbar^2\nabla^2}{2m} + \mu & 0\\
0 & \hbar \partial_\tau + \frac{\hbar^2\nabla^2}{2m} + \mu
\end{pmatrix}
  \delta({\bf x}-{\bf x}')\delta(\tau-\tau')
\end{multline}
and the self-energy is
\begin{align}
  \hbar{\bf\Sigma}({\bf x},\tau;{\bf x}',\tau') = \begin{pmatrix}
    0 & \Delta({\bf x},\tau) \\ \Delta^*({\bf x},\tau) & 0
  \end{pmatrix}\delta({\bf x}-{\bf x}')\delta(\tau-\tau').
\end{align}
The first term in Eq.~(\ref{eq:Seff}) corresponds to the thermodynamic potential of a noninteracting gas that is the first term in Eq.~(\ref{Omega}). The BCS Green's function defined in the previous section is obtained from the terms in Eq.~(\ref{eq:Seff}) that are quadratic in the pairing fields. They may be written as
\begin{equation}
    S^{(2)}[\Delta,\Delta^*] =
   -\hbar\sum_{{\bf k},n} \Delta^*({\bf k},i\omega_n)
   \frac{1}{2\hbar T^{\rm MB}({\bf k},i\omega_n)}
   \Delta({\bf k},i\omega_n).
\end{equation}
Neglecting all higher-order terms and performing the Gaussian functional integral yields the second term in Eq.~(\ref{Omega}), where the factor $T^{\rm 2B}(0)$ appears due to the normalization involved in the Hubbard-Stratonovich transformation.

By introducing the renormalized pairing field $\phi({\bf k},i\omega_n) = \Delta({\bf k},i\omega_n)/\sqrt{Z}$ with the renormalization factor
\begin{equation}
  Z = \frac{2}{\hbar}\left[\frac{\partial}{\partial i\omega_n}\frac{1}{T^{\rm MB}({\bf 0},\omega^m_{\bf 0})} \right]^{-1},
\end{equation}
where $\omega^m_{\bf 0}$ is the frequency of the molecular pole in the many-body $T$-matrix at zero momentum, this quadratic action can be recast into the action for an ideal Bose gas
\begin{equation}
    S^{(2)}[\phi,\phi^*] =
   \sum_{{\bf k},n} \phi^*({\bf k},i\omega_n)
   (-i\hbar\omega_n + \hbar\omega^m_{\bf k})
   \phi({\bf k},i\omega_n).
\end{equation}
Lowest-order corrections to NSR due to interactions between pairs are then obtained from the fourth-order term of the effective action. Writing it as $S^{(4)}[\phi,\phi^*] \equiv (\hbar\beta V/2) |\phi_0|^4 T_{\rm mm}$, where $\phi_0=\Delta_0/\sqrt{Z}$, we thereby find the molecule-molecule scattering length
\begin{figure}
\begin{center}
\includegraphics[width=\columnwidth,draft=false]{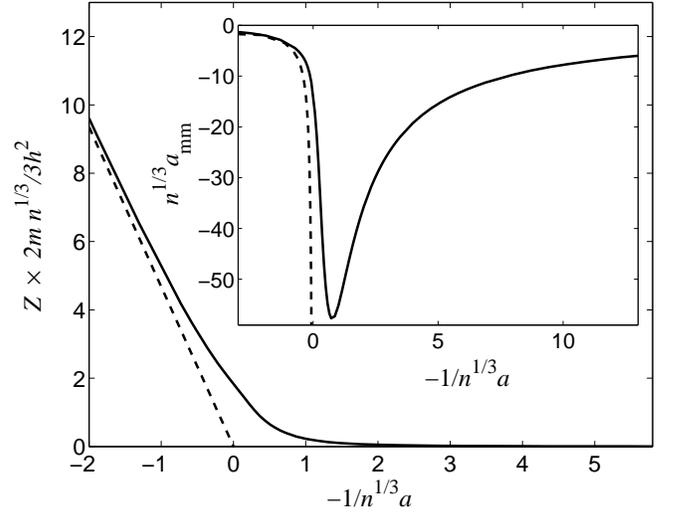}
\bf \caption{ \rm The renormalization factor $Z$ (solid line). Inset: the dimer-dimer scattering length $a_{\rm mm}$ (solid line). The respective BEC limits which are recovered as $a\rightarrow 0^+$ are shown as dashed lines.
\label{figZ}}
\end{center}
\end{figure}
\begin{multline}
  a_{\rm mm} =
  \frac{\sqrt{2m^5}}{\hbar^5}
  \frac{Z^2}{32\pi^3}
  \Bigg\lbrace\frac{-\pi}{(-4\mu)^{\frac{3}{2}}} -
  \\
  2\beta^2\int_{-\beta\mu}^\infty dx \frac{\sqrt{x/\beta+\mu}}{x^2}
   \frac{2}{e^x-1}\left[ \frac{1}{x} + 2 + \frac{2}{(e^x-1)}\right]
  \Bigg\rbrace.
  \label{eq:amm}
\end{multline}
The integrand in Eq.~(\ref{eq:amm}) is sharply peaked about $x=0$ thus in the BEC limit, where  $Z=16\pi\hbar^4/m^2 a$ and $\mu \simeq -\hbar^2/2m a^2$ is large and negative, the integral can be neglected and the molecule-molecule scattering length reduces to $a_{\rm mm} = -4 a$. Note that this is the same result found in Ref.~\cite{Romans04} up to a minus sign which was erroneously omitted there. Plots of $a_{\rm mm}$ and $Z$ are shown in Fig.~\ref{figZ}.

\begin{figure}
\begin{center}
\includegraphics[width=\columnwidth,draft=false]{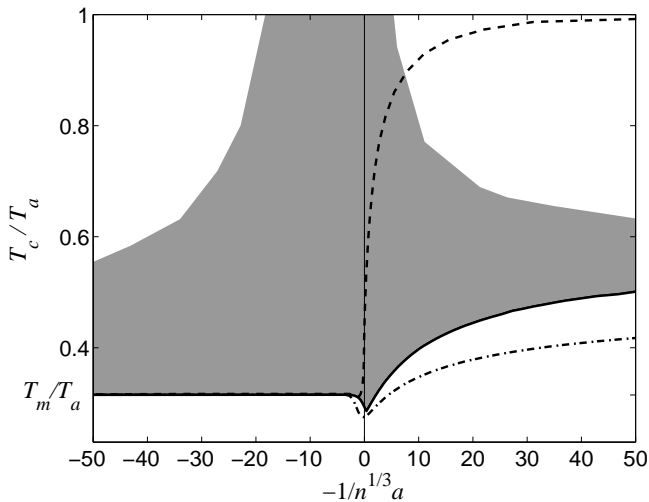}
\bf \caption{ \rm Universal phase diagram beyond NSR for the same parameters as in Fig.~\ref{fig1}. The solid line is $T_c$ computed including molecular interactions. The dashed-dotted and dashed lines are same many-body and two-body $T_c$ results respectively shown in Fig.~\ref{fig1}. The shading denotes the mechanical instability region.
\label{figTcs}}
\end{center}
\end{figure}

\begin{figure}
\begin{center}
\includegraphics[width=.92\columnwidth,draft=false]{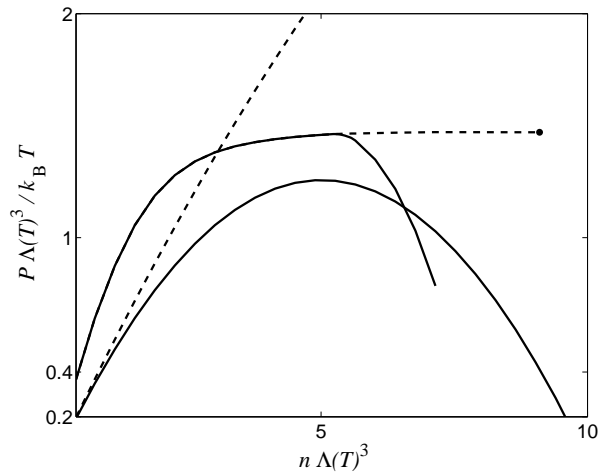}
\bf \caption{ \rm The pressure of the gas in the NSR formalism (dashed lines) and with molecular interactions beyond NSR included (solid lines). The curves beginning at $P\Lambda(T)^3/k_B T= 0.2$ and $0.4$ correspond to $-1/n^{1/3}a\simeq 10^{-3}$ and $10^{3}$, respectively.
\label{figP2}}
\end{center}
\end{figure}

The critical temperature obtained by including mean-field effects due to the stable dimers is plotted as the solid line in Fig.~\ref{figTcs}, and is always higher than the critical temperature predicted within the NSR formalism (dashed-dotted line) owing to the presence of an attractive interaction between the stable dimers throughout the crossover. The pressure of the gas for densities below the critical density is shown in Fig.~\ref{figP2} at various atomic scattering lengths (solid lines).
We see that the inclusion of molecule-molecule interactions gives rise to a region of negative compressibility, indicating a mechanical instability at temperatures above the critical temperature. The shading in Fig.~\ref{figTcs} denotes the resulting unstable region with an upper boundary given by the temperature for which $\partial n/\partial P = 0$. A striking feature of this phase diagram is the presence of a mechanically stable region at negative scattering lengths, where the gas remains in the normal phase far below the ideal gas critical temperature due to the many-body effects discussed above.

The effects of screening of the interatomic interaction may be estimated by dressing the atomic interaction in the many-body $T$-matrix with an RPA bubble sum,
\begin{equation}
T^{\rm MB}({\bf k},i\omega_n)=\frac{1}{[T^{\rm 2B}(z)]^{-1}-\Pi({\bf 0},0)-\Xi({\bf k},i\omega_n)},
\end{equation}
where
\begin{equation}
  \Pi({\bf 0},0) = -\frac{4}{\hbar^2\beta V}\sum_{{\bf k},n} G({\bf k},i\omega_n)^2
\end{equation}
and $G({\bf k},i\omega_n) = \hbar/(i\hbar\omega_n - \xi_{\bf k} )$. We have estimated that this correction raises the critical temperature by less than 15\% on the BCS side. The phase diagram obtained within a more refined calculation which includes screening fully self-consistently is beyond the scope of this paper but should therefore not deviate substantially from the one presented here.

\section{Cesium phase diagram}
\label{sec:Cs}
We now turn our attention to the specific case of a gas of $^{133}$Cs atoms near the narrow $s$-wave Feshbach resonance due to a $d$-wave molecule located at $B_0=47.78$~G~\cite{Lange09}.
\begin{figure}
\begin{center}
\includegraphics[width=\columnwidth]{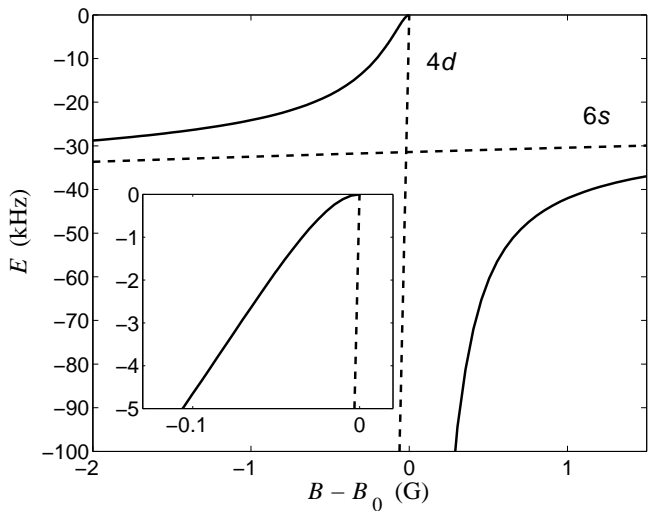}
\bf \caption{ \rm Binding energy of the $6s$ and $4d$ bare states (dashed lines) giving rise to the avoided crossing between the two dressed states (solid lines). In the close-up we show the Wigner threshold behavior of the binding energy near the continuum. \label{figBE}}
\end{center}
\end{figure}
The shallow $6s$ bound state where small inelastic losses were experimentally observed is present just below the atomic continuum and should be taken into account since it crosses the $4d$ state that causes the Feshbach resonance. Thus, the two-body $T$-matrix for the \textsuperscript{133}Cs $s$-wave Feshbach resonance considered here is obtained from Eq.~(\ref{T2B}) by replacing $a$ with $a_\mathrm{bg}(B)[1+\Delta \mu \Delta B/(z-\delta)]$~\cite{DuinePhD,Bruun05,Massignan08}. Here, the difference of magnetic moments between the $4d$ bound state and the atomic continuum is $\Delta\mu=1.15\mu_B$~\cite{Lange09}, and $a_\mathrm{bg}(B)$ is the background scattering length~\cite{endnote2}. The energy detuning from the Feshbach resonance is defined as $\delta = \Delta\mu(B-B_0)$ and the width of the resonance is found to be $\Delta B = 0.16$~G~\cite{Lange09}.

The energy of the poles of the cesium two-body $T$-matrix are plotted as solid lines in Fig.~\ref{figBE} which shows the avoided crossing between the $4d$ and $6s$ states. The binding energies of the bare states are shown as dashed lines. Also shown is a close-up near the Feshbach resonance showing the Wigner threshold behavior of the binding energy that is characteristic of an $s$-wave resonance. This mirrors the behavior of the binding energy in the two-body limit which is plotted as the dashed line in the inset of Fig.~\ref{fig2}. The many-body $T$-matrix is found by inserting the two-body $T$-matrix for cesium obtained above into Eq.~(\ref{TMB}).

\begin{figure}
\begin{center}
\includegraphics[width=\columnwidth]{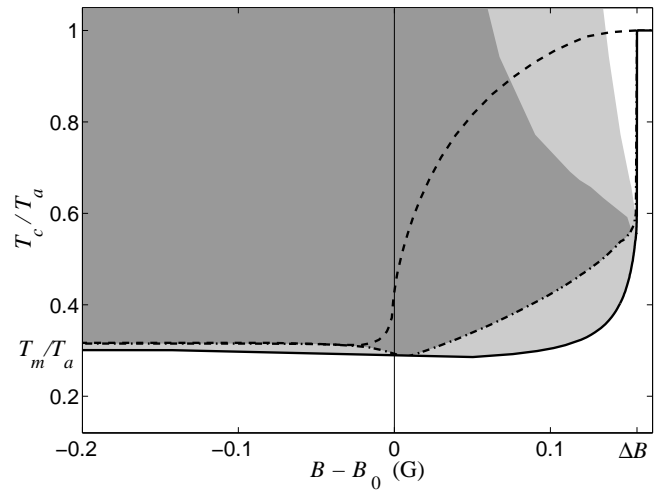}
\bf \caption{ \rm Critical temperature for pair condensation $T_c$ in a gas of \textsuperscript{133}Cs for particle densities $n=10^{11}\mathrm{cm}^{-3}$ (dash-dotted) and $10^{14}\mathrm{cm}^{-3}$ (solid), and the corresponding instability regions (dark and light grey, respectively).
For $n=10^{14}\mathrm{cm}^{-3}$ we have $T_a$=79 nK and $T_m$=25 nK. Note that the curve has no kink at $B-B_0=\Delta B=0.16G$ although this is not visible on this scale. The dashed line is the two-body result for $n=10^{11}\mathrm{cm}^{-3}$. \label{figCs}}
\end{center}
\end{figure}

The experimentally relevant crossover is between a BCS-like state in the atomic continuum and a gas of condensed dimers occupying the upper branch of the avoided crossing.
To describe theoretically this metastable configuration, we neglect the contribution to the thermodynamic potential that arises from the deeper-lying pole of the cesium two-body $T$-matrix corresponding to the lower branch of the avoided crossing. The resulting critical temperature for two densities is plotted in Fig.~\ref{figCs}.
As the magnetic field is increased across the Feshbach resonance, the character of the pairs changes gradually from stable and deeply-bound dimers to infinite-lifetime pairs weakly bound by many-body effects. The density of the pairs decreases gradually as $B$ is increased, reaching zero as $B-B_0$ reaches $\Delta B$, where the scattering length vanishes. For $B\geqslant B_0 + \Delta B$ there are no longer any pairs and both $T_c$ and $T_c^a$ reach their asymptotic value of $T_a$. In principle, experiments hint that \textsuperscript{133}Cs molecules may be sufficiently stable against decay caused by inelastic collisions~\cite{Ferlaino08,Knoop09} but we find that the crossover will nevertheless be subject to a mechanical instability due to the attractive interaction between the molecules. Inelastic losses can also be induced in the vicinity of the Feshbach resonance by the presence of Efimov states. Indeed, by using methods employed in Ref.~\cite{Massignan08}, we find that these three-body states lie in the narrow magnetic field range of $|B-B_0|\lesssim 0.01$~G.

The mechanical instability may be overcome on the BEC side by a positive background scattering length for the molecules, however it is not possible to properly treat this in a single-channel NSR setting and the inclusion of this and other effects of molecular interactions is an interesting problem for future exploration in the context of a two-channel model.

\section{Outlook and conclusions}
\label{sec:end}

The effects described in this paper may be readily investigated in a variety of ways.
Pairs of bosonic atoms can be experimentally detected using techniques that have been successfully employed to observe fermionic-atom pairs, for example, by the measurement of correlations in atom shot noise~\cite{Greiner05} or by radio-frequency spectroscopy~\cite{Chin04Ketterle}. Furthermore, the experimental observation of a pronounced reduction in the critical temperature for condensation of paired and unpaired atoms is in itself compelling evidence that pairing is taking place. As is shown in Figs.~\ref{figTcs} and~\ref{figCs}, this striking effect can indeed be seen at relatively high temperatures where the gas remains mechanically stable.

Although the mechanical instability of the cesium gas may be overcome on the BEC side of the resonance due to a positive background scattering length for the molecules as mentioned in the previous section, it remains to be shown experimentally whether the gas is stable enough against inelastic losses. Indeed, measurements performed in Ref.~\cite{Ferlaino08} found inelastic losses due to dimer-dimer collisions of $1\times10^{-11}$cm\textsuperscript{3}/s around $B=46$~G which then increase for a fixed temperature as $B_0$ is approached. On the other hand, the loss rate was found to be strongly temperature-dependent and this may serve to significantly stabilize the gas near $T_c$.

Although not the focus of this paper, the condensation of unpaired atoms can also be used to experimentally shed light on some other interesting physics. For example, below $T_c$ a transition between bound and unbound \mbox{(half-)vortex} pairs will take place as the $T_c^a$ line is crossed from the PC+AC phase to the PC phase~\cite{Radzihovsky04,Romans04}. It would be very exciting if this could be directly observed as in the case of the Kosterlitz-Thouless transition in a two-dimensional Bose gas~\cite{KTtransition}. However, it is not possible to make a definite prediction regarding the mechanical stability of the gas across this phase transition within the framework presented here~\cite{Yin08}.
%
%
Further insight may be obtained by using a two-channel approach to treat molecular interactions beyond mean-field theory. This has the added advantage that a background scattering length for the molecules as well as atom-molecule interactions can be readily included, and would be an interesting topic for further research.
\begin{acknowledgments}
The authors would like to thank Steven Knoop for fruitful discussions that inspired this project and for providing data on the Feshbach resonance. We also wish to thank Georg Bruun for his help in the early stages of the calculation. This work is supported by the Stichting voor Fundamenteel Onderzoek der Materie (FOM) and the Nederlandse Organisatie voor Wetenschaplijk Onderzoek (NWO), the Spanish MEC project TOQATA (FIS2008-00784) and the ESF/MEC project FERMIX (FIS2007-29996-E).
\end{acknowledgments}

\begin{thebibliography}{25}
\bibitem{endnote2}
  \bibinfo{note}{
  The background scattering length is magnetic field-dependent predominantly due to a
  Feshbach
  resonance at $-11.74$G which is unrelated to the Feshbach resonance of interest
  here. It behaves as
  $a_\mathrm {bg}(B)/a_0=(1722+1.52B)[1-28.72/(B+11.74)]$~\cite{Kraemer06}.}
\end{thebibliography}

\end{document}